# Beyond Decoherence: Control the Collective Quantum Dynamics of Quasi Particles in Topological Interface

*Fatemeh Davoodi*[1,2,*]

[1]*Nanoscale Magnetic Materials, Institute of Materials Science, Kiel University, 24143, Kiel, Germany*

[2]*Kiel Nano, Surface and Interface Science KiNSIS, Christian Albrechts University, Kiel, Germany*

[*]*fda@tf.uni-kiel.de*

**Abstract:** Long-lived coherent quasiparticles are a promising foundation for novel quantum technologies, where maintaining quantum coherence is crucial. Decoherence, driven by finite emitter lifetimes, remains a central challenge in quantum computing. Here, we control the dynamics of spatially separated quantum emitters via preserving their phase information by introducing a topological waveguide as a robust chiral reservoir. Incoherent quantum emitters randomly positioned near a perturbed honeycomb plasmonic interface and couple to the mutual topological interface mode. Using the $S_3$ Stokes parameter, we trace far-field polarization patterns that reflect emitter coherence and spin–momentum locking. We show that even weakly coupled emitters exhibit coherent excitation and imprint phase on the emission. Time-domain dynamics reveal signatures of superradiance and subradiance that correlate with spatial interference in $S_3$. These spatial-temporal features confirm that the observed polarization patterns arise from coherent quantum many-body dynamics, not classical interference. This challenges the conventional dichotomy between incoherent and coherent regimes, revealing that topological chiral photonic environments mediate long-range quantum correlations beyond standard waveguide QED.

**Introduction:**

Quantum coherence, the ability of a quantum system to maintain phase relations between different states, is the foundation of quantum technologies[1–5], from entanglement-based communication[4,6–9] to fault-tolerant quantum computation[10–12]. However, in realistic systems, interactions with the environment lead to [13], the rapid degradation of coherence through loss of phase information. This transition from quantum to classical behavior, driven by environmental coupling, is a central challenge in condensed matter and quantum information science.[14–17] In solid-state platforms, quasiparticles such as excitons[18–26] plasmons-polaritons[22,24,26–28] are especially vulnerable to decoherence due to finite lifetimes, scattering, and interaction with vibrational modes. A critical milestone in overcoming this challenge lies in engineering environments that preserve coherence rather than destroy it. One promising strategy involves embedding quantum emitters into structured photonic environments that suppress decoherence by symmetry[29–33], topology[34–39], or interference[40–43]. In particular, topological photonic interfaces offer a fundamentally new paradigm. These systems support edge states that are robust against disorder, exhibit directional transport, and enforce spin–momentum locking, making them ideal candidates for mediating protected quantum interactions.[44–49] In this work, we introduce a platform that leverages a topologically protected chiral plasmonic waveguide to preserve and mediate quantum coherence between spatially separated emitters, even in the weak coupling regime. While conventional waveguide quantum electrodynamics (QED) requires strong coupling to realize coherent emitter interactions[50–53], our system demonstrates that long-range coherence and collective emission can persist when quantum emitters are only weakly coupled to the bath, an outcome made possible by the directional, phase-preserving, and disorder-resilient nature of the chiral edge states. This represents a novel coupling regime in open quantum systems: one where the environment not only fails to decohere the system but actively protects and channels phase-coherent interactions across mesoscopic distances. We develop a cascaded Lindblad framework that explicitly incorporates directional reabsorption, topological robustness, and local dephasing superoperators, and show that coherence can be stabilized well beyond the near-field and strong-coupling limits. Our approach connects key threads in modern nanophotonics: quasiparticle coherence, hybrid light–matter coupling[54] (as in plasmon–exciton systems[22]), and topological band engineering. The result is a new class of non-reciprocal, coherence-preserving quantum interfaces that overcome traditional decoherence constraints and open new pathways for scalable, robust quantum photonic technologies. Topological photonic insulators are artificially engineered materials that mimic the quantum spin Hall effect (QSHE) in electronic systems, lie at the heart of this approach.[46] These systems preserve a form of time-reversal symmetry (TRS) and support spin-momentum–locked edge states, where the polarization or angular momentum of light plays the role of pseudospin. [46,49,55] Such helical photonic states, described by Kramers-like pairs, exhibit unidirectional transport and immunity to backscattering from imperfections—properties inherited from their nontrivial topological band structure. Topological order thus not only redefines the classification of quantum phases but also introduces fundamentally robust building blocks for quantum photonics.[56] The realization that photonic systems can host topological states has led to a surge in experimental platforms—ranging from photonic crystals and coupled resonators to metamaterials and polaritonic lattices.[55] A particularly promising avenue is the integration of quantum emitters (QEs) into topological photonic environments, enabling the suppression of decoherence via topologically protected light–matter interactions.

In our work, we explore precisely this regime: the coupling of gold nanoparticles modeled as two-level quantum emitters to the helical edge states of a one-dimensional (1D) topological photonic interface. This interface arises from a zigzag domain wall between two distinct phases of a two-dimensional plasmonic crystal, formed by perforating a gold film with a honeycomb lattice of subwavelength triangular nanoholes (see Fig. 1a). By carefully choosing 60 nm gold nanospheres, the dipole resonance of each emitter is spectrally aligned with the topological bandgap (shaded gray region in Fig. 1h,i), ensuring strong interaction with the chiral edge modes. This configuration allows us to probe how quantum coherence and collective emission evolve as we vary the emitter–bath coupling strength. In particular, we demonstrate that even in the weak coupling regime, where conventional models predict rapid decoherence, the chiral edge mode mediates non-trivial long-range quantum correlations, made possible by topological protection, non-reciprocal energy transfer, and phase robustness. Our system thus provides a tunable platform to investigate the crossover from Markovian to non-Markovian quantum dynamics, and to realize scalable, directional entanglement protocols in photonic nanostructures.

We design a two-dimensional plasmonic crystal consisting of a honeycomb lattice of subwavelength triangular nanoholes perforated in a gold film. The unit cell is a hexagonal cluster of six triangular holes, the lattice constant of a, with $a_1$ and $a_2$ as the lattice vectors, arranged in a honeycomb pattern (Fig. 1b-c). In the symmetric (unperturbed) configuration (Fig.1b), this structure possesses a Dirac-like band degeneracy at the center of the Brillouin zone (Γ point): specifically, two doubly-degenerate "Dirac cones" overlap in energy, producing a four-fold degeneracy at Γ point. This degeneracy is protected by the six-fold rotational ($C_6$) symmetry of the hexagonal hole cluster. Each degenerate mode can be viewed as an orbital-like state (e.g. s, p, d orbital analogues) with a two-fold pseudospin degeneracy (stemming from two opposite circulating current directions).[57,58] By introducing a controlled inversion-symmetry-breaking deformation to the nanohole cluster, a bandgap opens at the Dirac point (Fig.1d,e) We achieve this by moving the holes either inward (shrinking the hexagon) or outward (expanding the hexagon) within each unit cell. These two deformation choices produce opposite band orderings, i. one topologically trivial and ii. one inverted, analogous to the sign change of the mass term in the Bernevig–Hughes–Zhang (BHZ) model for the QSHE.[59]

**Results**

**Theoretical Model**:

To model collective plasmonic modes in the perforated gold film, we use a coupled dipole (multipole) approach in the quasi-static regime.[59,60] Each triangular nanohole supports a localized surface plasmon (LSP) mode, modeled as an out-of-plane electric dipole due to the dominance of transverse-magnetic (TM) polarization in the topological bandgap.[57] Assuming linear and nonmagnetic media, the induced dipole moment $p_i$ of the nanoholes centrally located at $r_i$ is linearly proportional to the electric field at $r_i$. In the absence of external excitations, this electric field is sum of the contributions of all other dipole moments in the array:[60]

$$p_i = \alpha_{eff}(\omega) \sum_{i \neq j} \mathbf{G}(r_{ij}, \omega) p_j \qquad (1)$$

We assign an effective polarizability tensor $\alpha_{eff}(\omega)$ to each hole's fundamental mode, $\mathbf{G}(r_{ij}, \omega)$ is the dyadic Green's function describing dipole–dipole interactions. In free space, this tensor takes the form:[61]

$$G(r) = \frac{1}{4\pi\varepsilon_0}\left[\left(1 + \frac{ik_0 r - 1}{k_0^2 r^2}\right)\mathbf{I} + \left(\frac{3 - 3ik_0 r - k_0^2 r^2}{k_0^2 r^2}\right)\hat{r}\hat{r}\right] e^{ikr}/r \qquad (2)$$

where $k_0 = \omega/c$ is the free space wavenumber, $I$ is the identity matrix, and $\hat{r}$ is the unit vector along $r$.

$\alpha_{eff}(\omega)$, the effective polarizability is derived using $\alpha_{eff}(\omega) = V\frac{\varepsilon(\omega) - \varepsilon_0}{\varepsilon_0 + L_z(\varepsilon(\omega) - \varepsilon_0)}$, $V = \frac{\sqrt{3}}{4}s^2 h$ for an equilateral triangular prism of side $s$ and height $h$, and the depolarization factor $L_z \approx \frac{h}{2s}$. This interaction formalism naturally maps onto a tight-binding (TB) model for collective plasmon modes on a 2D honeycomb lattice, where LSPs serve as orbitals and Green's function-mediated dipole interactions define hopping amplitudes. The resulting eigenvalue problem is:

$$\frac{1}{\tilde{\alpha}(\omega)} p_i = \sum_{i \neq j} G_\perp(r_i - r_j) p_j \tag{3}$$

which, in matrix form, becomes Hp = $\lambda$p, with $\lambda = \frac{1}{\tilde{\alpha}(\omega)}$ acting an eigenvalue and interaction Hamiltonian $H$.

Each triangle in the honeycomb lattice supports a resonant mode. Label the triangles such that the unit cell contains three sites on sublattice $A$ and three on sublattice $B$, arranged alternately around a hexagon. We denote these basis states as $|A_1\rangle$, $|A_2\rangle$, $|A_3\rangle$; $|B_1\rangle$, $|B_2\rangle$, $|B_3\rangle$, where $A_i$ and $B_i$ are opposite each other across the center of the hexagon (with $i = 1,2,3$ cyclically). The tight-binding Hamiltonian on a honeycomb lattice considering both NN ($t_1$) and NNN interactions ($t_2$) based on $c_i^\dagger$ and $c_j$ are the creation and annihilation operators at sites $i$ and $j$.

$$H = -t_1 \sum_{\langle i,j \rangle} c_i^\dagger c_j - t_2 \sum_{\langle\langle i,j \rangle\rangle} c_i^\dagger c_j + H.c. \tag{4}$$

In momentum space, this forms a block matrix:

$$H = \begin{bmatrix} H_{AA} & H_{AB} \\ H_{BA} & H_{BB} \end{bmatrix} \tag{5}$$

where $H_{AA}$ and $H_{BB}$ describe same-sublattice (NNN) interactions and $H_{AB} = H_{BA}^\dagger$ describes NN. We can write $H_{AB}$ as:

$$H_{AB} = t_1 \begin{pmatrix} 1 & 0 & e^{-i\mathbf{k}\cdot\delta_3} \\ e^{-i\mathbf{k}\cdot\delta_1} & 1 & 0 \\ 0 & e^{-i\mathbf{k}\cdot\delta_2} & 1 \end{pmatrix} \tag{6}$$

Here $\delta_{1,2,3}$ are the relative position vectors connecting an $A$ site to its three $B$ neighbors. The phase factors $e^{-i\mathbf{k}\cdot\delta_j}$ encode the Bloch boundary conditions for hopping to a neighboring cell in the $j$th direction.

This degeneracy can be lifted by breaking inversion symmetry of the unit cell. We incorporate this by splitting the nearest-neighbor coupling: intra-cell coupling $t_{intra}$ vs. inter-cell coupling $t_{inter}$. In the "expanded" case, (Fig.1 d), $t_{intra} < t_{inter}$, whereas in the "shrunken" case, (Fig.1 h), $t_{intra} > t_{inter}$. We map the system onto a two-orbital TB Hamiltonian analogous to the BHZ model[58,59]. In an appropriate basis $(|p_\pm\rangle = |p_x\rangle \pm$ $i|p_y\rangle$, $|d_\pm\rangle = |d_{xy}\rangle \pm i|d_{x^2-y^2}\rangle$ for pseudospin $\pm$), (See (Fig. 1f)) the Bloch Hamiltonian near $\Gamma$ can be linearized as $H_{\text{lin}}(\mathbf{k}) \approx v\,(k_x\Gamma_1 + k_y\Gamma_2) + m\Gamma_3$, where $\Gamma_i$ are 4×4 Dirac matrices and $m$ is the mass term proportional to $(t_{\text{intra}} - t_{\text{inter}})$. This gives rise to a photonic bandgap with topological edge states characterized by a nonzero Berry curvature $\Omega(\mathbf{K})$ at each momentum point $\mathbf{K} = (k_x, k_y)$ and Chern number:[47]

$$C = \frac{1}{2\pi} \iint \Omega(\mathbf{K}) d^2k \tag{7}$$

To implement this in a realistic platform, we design a honeycomb topological plasmonic lattice composed of perforated gold films. The unit cell has a lattice constant $a$=471.9 nm, with equilateral triangular apertures of side length $S$=148.46 nm patterned into a 40 nm-thick gold slab. This geometry supports a photonic double Dirac cone at the $\Gamma$-point, yielding a fourfold degeneracy at approximately 1.57 eV (Fig. 1g). By varying the deformation parameter ($a/R$), we realize the necessary symmetry-breaking conditions to open a topological bandgap and induce edge-localized chiral plasmonic modes (Fig. 1h,i). The mode anticrossing behavior observed in Fig. 2a occurs near the critical symmetric value ($a/R \approx 3.0$, corresponding to), evidencing the transition point between topologically trivial and nontrivial phases. This avoided crossing arises as a direct consequence of broken inversion symmetry. Insets in Fig. 2a depict the simulated electric field distributions on either side of this transition, highlighting the inversion of parity and orbital character between modes, a hallmark of band inversion.

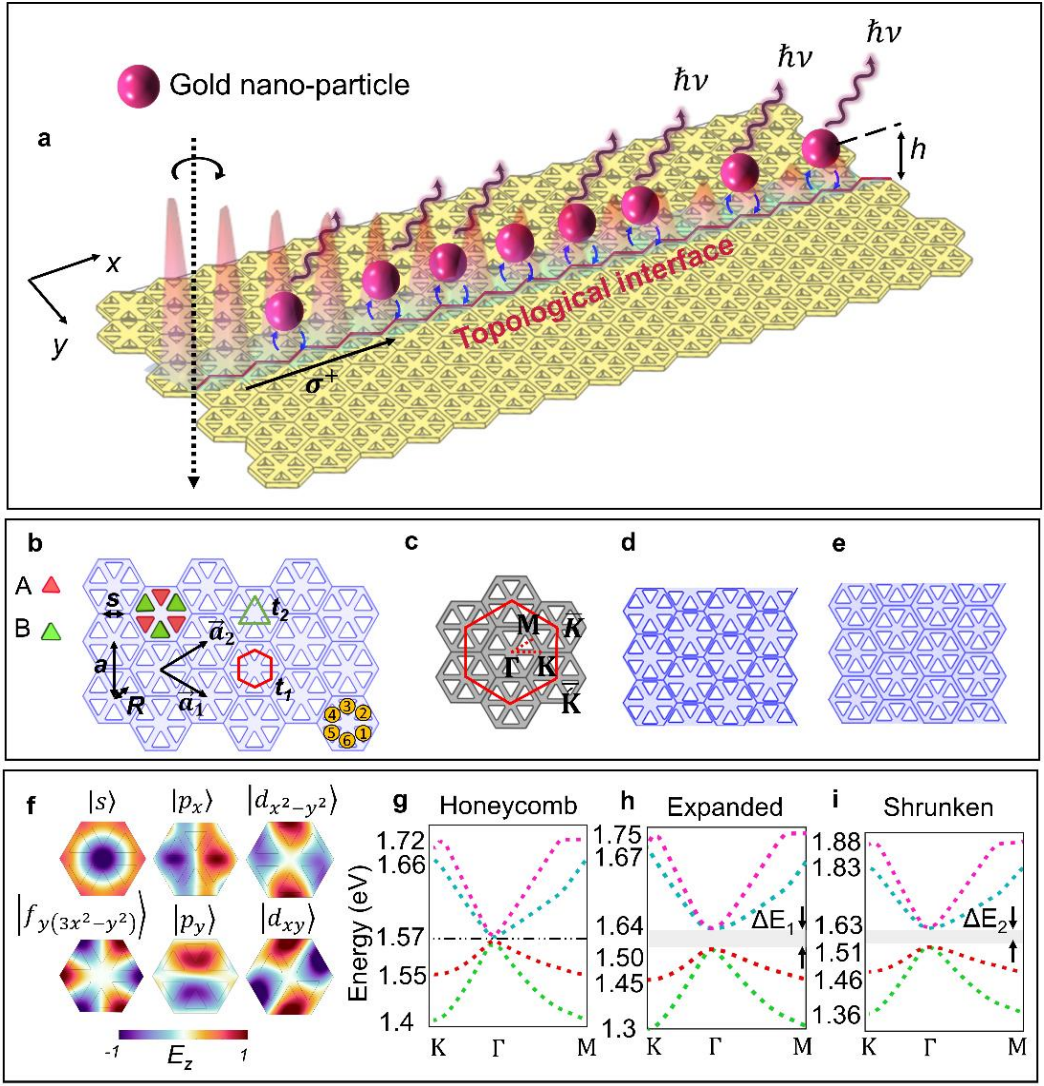

Figure 1: Topological plasmonic crystal design and band structure. **a**. Real-space layout of the honeycomb lattice with unit cell size of a=471.9 nm contracted by cluster of six perforated triangular nanoholes with edge size of s=148.46nm in a 40nm-gold film (red and green triangles denote two sublattices A and B). Six holes form a hexagonal cluster constituting the unit cell; a is the lattice constant and $a_{1,2}$ are lattice vectors. Black arrows indicate coupling: nearest-neighbor coupling $t_1$ (between A–B holes) and next-nearest neighbor $t_2$ (between holes of the same sublattice). **b**. Brillouin zone of the honeycomb lattice (red hexagon) with high-symmetry points Γ, K, M indicated. **c, d.** Two deformed lattice configurations: **c**. "expanded" lattice (outward hole shift) and **d**. "shrunken" lattice (inward hole shift), which have inverted band orderings relative to each other. **e**. Representative mode profiles (out-of-plane electric field $E_z$) of the six localized plasmonic modes in the hexagonal unit cell, labeled by their orbital analogues $|s\rangle$, $|p_x\rangle$, $|p_y\rangle$, $|d_{x^2-y^2}\rangle$, $|d_{xy}\rangle$, and $|f_{y(3x^2-y^2)}\rangle$. Each mode exists as a Kramers pair (pseudo-spins $\sigma_\pm$ corresponding to clockwise vs. counter-clockwise field rotation). **f**. Photonic band structure for the symmetric honeycomb lattice, showing a four-fold degenerate double Dirac cone at Γ (shaded gray region). Two bands (magenta, cyan) correspond to the p-orbital doublet and two (red, green) to the d-orbital doublet, crossing at the Dirac frequency ~1.57 eV. **g,h**. Band structures for the expanded **g** and shrunken **h** lattices. A bandgap opens at Γ in both cases (gap highlighted in gray), but the ordering of p (cyan/magenta) vs. d (green/red) bands is inverted between **g** and **h**. $\Delta E_1$, $\Delta E_2$ mark the differing gap edge positions in the two cases, signifying a topological band inversion.

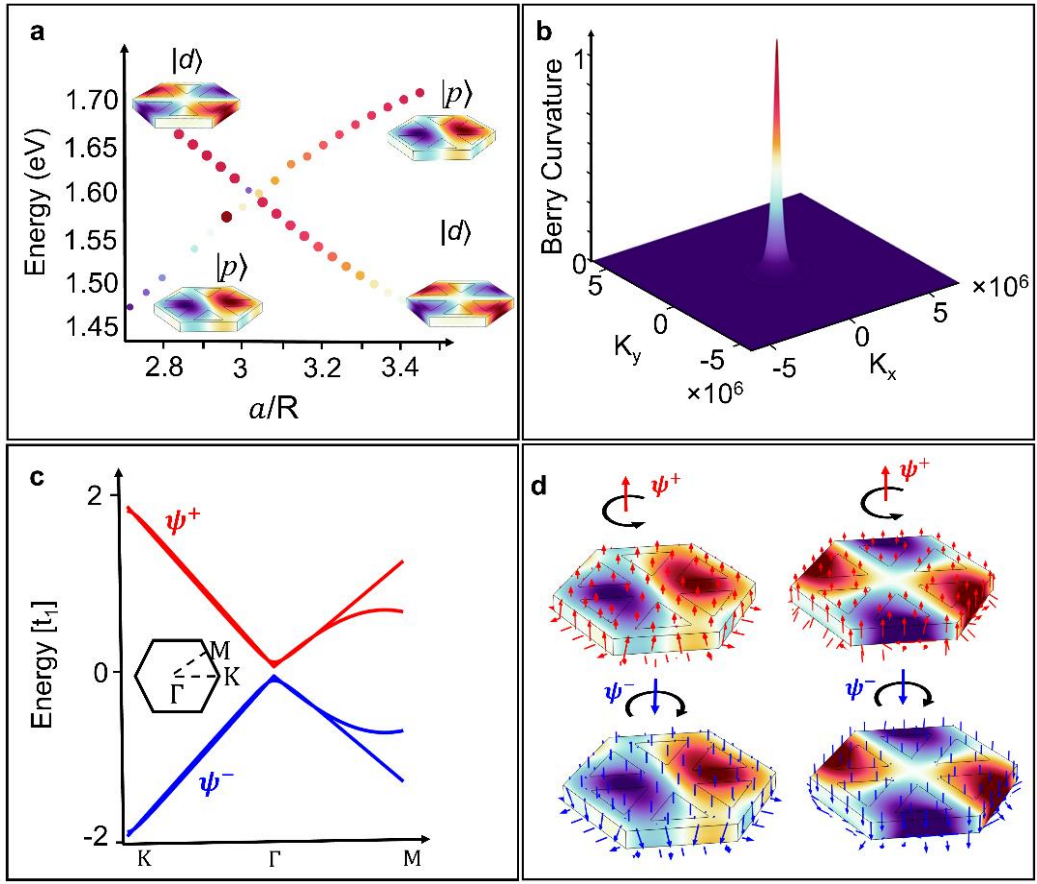

Figure 2: Topological band inversion and Berry curvature. **a**. Evolution of two representative band eigenfrequencies (points) as a function of the structural deformation parameter (hole spacing a relative to hole size R). At the critical value (around a/R ≈ 3.0), the bands cross and exchange character, indicating a topological transition. Insets: out-of-plane field profiles of the mode before (left) and after (right) the inversion, corresponding to predominantly d-like vs. p-like character, respectively. **b**. Computed Berry curvature (for one pseudospin sector) in k-space for the topologically nontrivial band, showing a pronounced peak at the Brillouin zone center (Γ point), whose integral yields the spin Chern number $C_s$= 1. **c**. Tight-binding band structure from the effective BHZ model, capturing the inverted band ordering. Red and blue curves correspond to the two pseudospin sectors ($\psi^+$ and $\psi^-$), which are degenerate in energy but carry opposite spin. (Inset: 2D hexagonal Brillouin zone with high-symmetry points.). **d**. Pseudospin-resolved modal fields in the unit cell for the topologically nontrivial case. The top row shows the $\psi^+$ mode (red, with upward circular electric dipole moment indicated), and the bottom row shows $\psi^-$ (blue, with downward oriented dipole). The arrows on the hexagonal clusters depict the local OAM of the mode field, demonstrating opposite handedness for the two pseudospins. This optical spin distinguishes the two degenerate modes and plays the role of an effective spin degree of freedom, leading to spin-momentum locking in the edge transport.

Correspondingly, Fig. 2b shows that the Berry curvature is strongly localized near the Brillouin zone center in the topological regime, confirming the emergence of topological order. To further quantify the topological phase, we calculate the spin Chern number $C_s$= (C↑ – C↓)/2 = 1, indicative of a $Z_2$ topological photonic insulator. This nontrivial topology is corroborated by spin-resolved Berry curvature calculations, where the two pseudospin channels exhibit sharply peaked but oppositely signed Berry curvature near the Γ-point. This is consistent with a band-inverted regime supporting a photonic quantum spin Hall effect.[46] The eigenstates $|p_\pm\rangle$ and $|d_\pm\rangle$, which serve as a suitable basis, carry well-defined orbital angular momentum (OAM) and form the backbone of the topological interface modes. Importantly, the two chiral edge states that emerge within the bandgap represented as ψ⁺ and ψ⁻ in Fig. 2c,d are degenerate in energy but carry opposite angular momentum. These results are validated using full-wave COMSOL simulations, which not only capture the spatial mode distributions but also reflect the OAM-based helicity and spin-momentum locking of the chiral edge plasmons. This provides a rigorous and consistent link between the tight-binding theory, multipole model, and photonic topological band structure observed in the numerical design.

**Helical Edge States and Spin–Momentum Locking:**

A defining consequence of the bulk band topology is the emergence of helical edge states at the interface between two photonic domains of different topology. We construct a domain wall by adjoining a topologically nontrivial region (expanded lattice) next to a trivial region (shrunken lattice). Within the common bandgap (~1.55–1.65 eV in Fig. 1h, i). These are the photonic analogs of QSHE edge modes: they come in counter-propagating pairs with opposite pseudospin (circular polarization). In other words, one helical mode (ψ⁺) is predominantly right-circularly polarized and propagates, say, to the right, while its time-reversed partner (ψ⁻) is left-circularly polarized and propagates to the left. This optical spin–momentum locking is enforced by the underlying pseudospin conservation and TRS. Backscattering is strongly suppressed because there is no way to reverse direction without flipping the polarization (spin) of the mode, which a TR-symmetric perturbation cannot induce. Thus, ideally, these edge modes propagate unidirectionally and are immune to localization from imperfections as long as the perturbations do not break the pseudospin symmetry. This robust transport was confirmed in full-field simulations: an edge mode launched at one end travels around sharp bends with negligible reflection. Fig. 3a,b, illustrate the localization and one-way transport of the edge mode. In Fig. 3a, the simulated field profile (out-of-plane $E_z$ component at 1.6 eV) is concentrated along the domain wall and decays into the bulk on either side. Fig. 3b shows a snapshot of a wavepacket launched along the interface; the energy stays confined to the interface as it travels. In contrast, if the lattice is uniformly trivial (or the excitation frequency lies outside the bandgap), no guided edge transport is observed, the field either radiates into the bulk or remains localized near the source (Fig. 3b). The band dispersion of the helical edge states is shown in Fig. 3c: within the bulk bandgap (shaded region), a pair of counter-propagating linear dispersions appears (red vs. blue curve). These correspond to the $\sigma^+$ and $\sigma^-$ edge modes traveling in opposite directions. Notably, each mode is polarized with a definite helicity locked to its direction of motion (indicated by red and blue circular arrows).

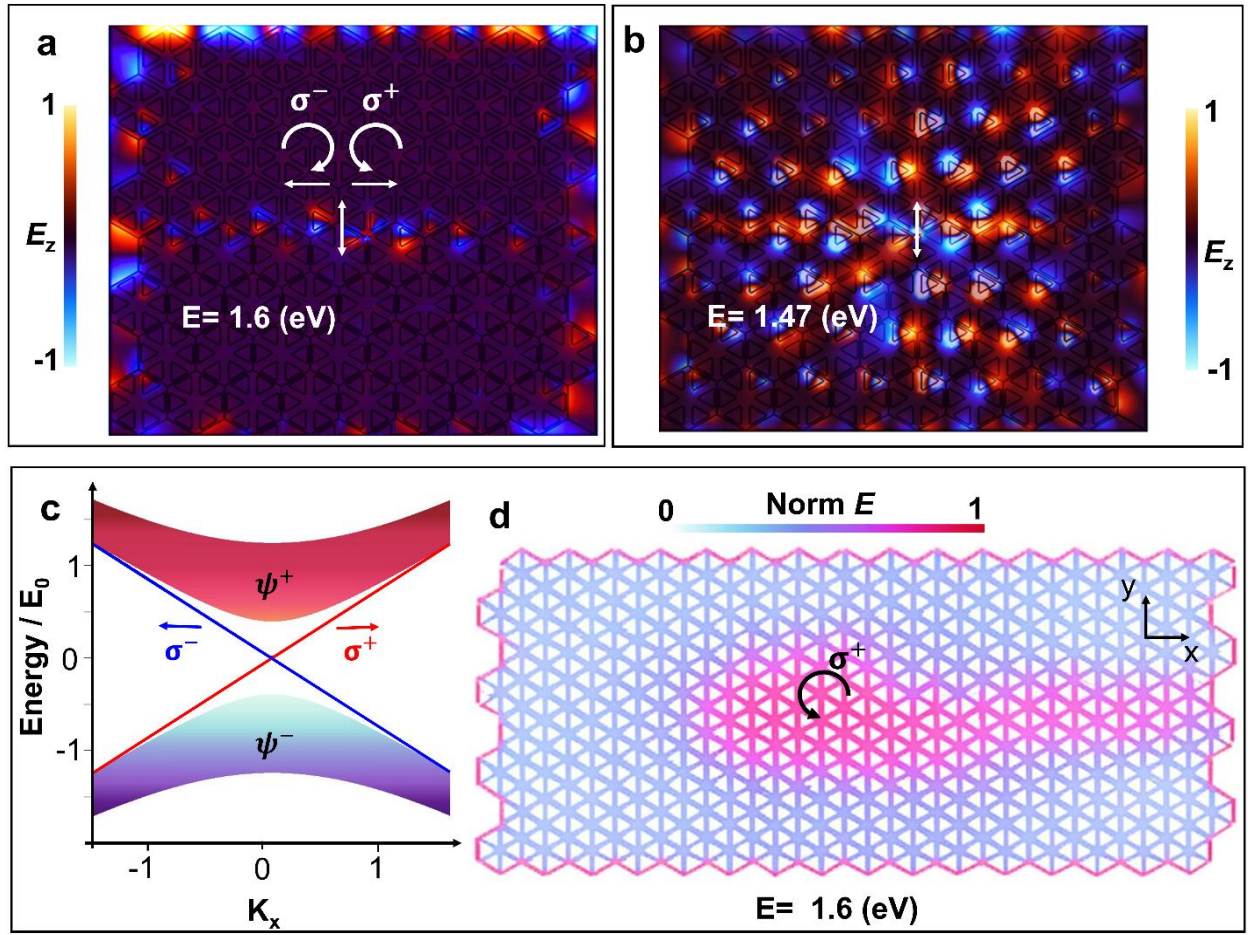

Figure 3: Helical topological edge modes at a domain wall. **a**. Simulated out-of-plane electric field ($E_z$) profile of an edge eigenmode at E = 1.6 eV, localized at the interface between a topological (expanded) and trivial (shrunken) domain (interface runs horizontally in the middle). The mode is confined along the edge and decays into both bulk regions. **b.** Real-space steady-state field when a source excites the topological edge mode from the left end of the interface (white arrows denote direction of energy flow). The mode propagates to the right without loss, even around corners, illustrating robust one-way transport. **c**. Field pattern at E = 1.37 eV in a uniform trivial lattice (no topological interface). The excitation remains localized near the source and radiates isotropically into the bulk, indicating the absence of edge transport in the trivial phase. **d**. Calculated band structure of edge states (red for $\psi^+$, blue for $\psi^-$) within the bulk bandgap. The two pseudospin channels cross linearly at the interface (Dirac-like dispersion) and carry opposite group velocities

(directions indicated by arrows). **e.** Schematic of optical spin–momentum locking at the interface. A σ⁺ polarized dipole (black dipole icon) couples into the right-moving mode (σ⁺ state, red arrow along edge) and not into the left-moving mode (σ⁻, blue arrow), enabling unidirectional, chiral excitation of the waveguide.

**Long-Range Chiral Emitter Interactions:**

Gold spherical nanoparticles (NPs) with a radius of 60 nm serve as plasmonic resonators exhibiting dipolar localized surface plasmon resonances. When positioned near the topological interface, these nanoparticles act as quantum emitters that couple to the helical edge states and to one another through the chiral photonic environment. Each NP is modeled as a quantum two-level emitter with a single optical transition $|g\rangle \leftrightarrow |e\rangle$, characterized by a dipole transition frequency $\omega_a$ (E_a~1.6 eV) tuned to the topological frequency $\omega_T$ (E_T~1.6 eV) which lay in the topological bandgap. This spectral alignment enables non-reciprocal light–matter interactions, wherein the coupling of the emitter to guided versus non-guided photonic modes becomes direction-dependent. The directional scattering is quantified by the complex coupling coefficients $\beta_\pm = \frac{\gamma_\pm}{\gamma_+ + \gamma_- + \xi}$, where $\gamma_\pm$ is the ratio of the spontaneous emission rate into σ± states.[62,63] $\Gamma_{wg} = \gamma_+ + \gamma_-$ is decay rate into waveguide modes, $\xi$ is the emission rate into bulk modes. Due to the excitation of one of the σ± states, $\beta_\pm = 1$ and $\beta_\mp = 0$, enforcing unidirectional emission and enabling coherent energy transfer mediated by the topologically protected interface. This leads to unconventional quantum optical effects, directional bound states (an emitter in a topological bandgap binds a photon that decays only to one side) and cooperative emission that depends on the topology of the photonic structure. The forward and backward emission rates are $\gamma_\pm \propto |\mathbf{d}^* \cdot \boldsymbol{\varepsilon}_\pm|^2$, where $\mathbf{d}$ and $\boldsymbol{\varepsilon}_\pm$ are complex transition dipole matrix element and the complex electric field amplitude of the forward and backward mode, respectively, and the asterisk indicates complex conjugation.[63] One can write the total Hamiltonian, $H = H_{em} + H_T + H_I$, consist of the emitter Hamiltonian $H_{em} = \hbar\omega_a \sigma^+ \sigma^-$, $\omega_a$ is transition frequency of the emitter, $\sigma^+ = |e\rangle\langle g|$ and $\sigma^- = |g\rangle\langle e|$ are raising and lowering operators, the tight-binding Hamiltonian $H_T$, and the interaction Hamiltonian $H_I = -\mathbf{d} \cdot \mathbf{E}(r_{em})$ which describes Coupling between the quantum emitter and the local electric field. Using Markov approximations, Lindblad master equation for a single two-level emitter coupled to a lossy environment by:[14]

$$\dot{\rho} = -\frac{i}{\hbar}[H_{em} + H_I, \rho] + 2\Gamma \mathcal{D}[\sigma^-]\rho \tag{8}$$

where $\mathcal{D}[\sigma^-] = (2\sigma^-\rho\sigma^+ - \sigma^+\sigma^-\rho - \rho\sigma^+\sigma^-)$ is dissipator, for a single emitter, the Hamiltonian is affected by chirality, the pattern of emitted radiation inherent the chirality of photonic bath of the non-trivial topological interference mode (chiral reservoir), we look this behavior at far field and discuss it later. For example, in the interaction of the quantum emitter with $\sigma^+$ of topological interference, $H_I = -\mathbf{d} \cdot \mathbf{E}_{\sigma^+}(r_{em})$. When two or more two-level emitters are placed at the same distance $h$ above the interface and coupled via the topological interface, the model can become an attractive in quantum simulation because long-range interactions are the source of nontrivial many-body phases and dynamics, and are also very hard to treat classically. In the case of coupling quantum emitters couple into non-chiral modes, each emitter can emit into both forward and backward modes. A photon emitted by emitter A can propagate rightward and excite emitter B, and similarly a photon from B can propagate leftward to A. This two-way exchange leads to a symmetric "flip-flop" interaction: the two excitations can resonantly hop back and forth between the emitters. The effective Hamiltonian for two identical emitters in this regime has an exchange term of the form of $H_{ex} = \hbar J(d)\left(\sigma_+^{(1)}\sigma_-^{(2)} + \sigma_+^{(2)}\sigma_-^{(1)}\right)$. $d$ is the separation along the interface and $J(d) \propto \frac{\Gamma_{wg}}{2}\sin(k_0 d)$, $k_0$ is mode wavenumber. $\sin(k_0 d)$ dependence of the effective Hamiltonian is the hallmark of the coherent flip-flop interaction. So the master equation for two emitters coupling to nonchiral waveguide can be written as[14]

$$\dot{\rho} = -\frac{i}{\hbar}[H_{em} + H_{ex}, \rho] + \frac{(\Gamma_{wg}+\xi)}{2}\sum_{i=1}^{2}\mathcal{D}[\sigma_i^-]\rho + \frac{\Gamma_{wg}}{2}\left(\mathcal{D}[\sigma_1^- + e^{ik_0 d}\sigma_2^-]\rho + \mathcal{D}[\sigma_1^- + e^{-ik_0 d}\sigma_2^-]\rho\right) \quad (9)$$

where $\mathcal{D}[L]\rho \equiv 2L\rho L^\dagger - L^\dagger L\rho - \rho L^\dagger L$ is the Lindblad dissipator. In this case both superradiance and subradiance will appear due to symmetric and anti-symmetric combinations, and the joint emissions are explicitly distance-dependent which more complicated in $N$ quantum emitters case:

$$\dot{\rho} = -\frac{i}{\hbar}[H, \rho] + \sum_{i,j=1}^{N}\left(\Gamma_{wg}\cos(k_0|x_i - x_j|)\left(2\sigma_-^{(j)}\rho\sigma_+^{(i)}\left\{\sigma_+^{(i)}\sigma_-^{(j)}, \rho\right\}\right)\right) \quad (10)$$

the Hamiltonian $H$ includes coherent emitter-emitter interactions as $H = \sum_{i\neq j}\frac{\Gamma_{wg}}{2}\sin(k_0|x_i - x_j|)$, and density matrix of the entire system is in dimension of $(2^N \times 2^N)$. In the case of coupling quantum emitters couple into topological mode bath, it leads to the simplest master equation for a cascaded system:

$$\dot{\rho} = -i(H_{eff}\rho - \rho H_{eff}^\dagger) + \sum_{i<j} 2\gamma_{ij}\,\sigma_-^{(j)}\rho\sigma_+^{(i)} + \mathcal{L}_\varphi[\rho] \quad (11)$$

where $H_{eff} = H_{sys} - i\sum_{j=1}^{2N}\frac{\gamma_j}{2}\sigma_+^{(j)}\sigma_-^{(j)} - i\sum_{i<j}\gamma_{ij}\sigma_+^{(j)}\sigma_-^{(j)}$ and $\mathcal{L}_\varphi[\rho] = \sum_j \gamma_\phi\left(\sigma_z^{(j)}\rho\sigma_z^{(j)} - \rho\right)$ is pure dephasing which model the decoherence imposed in the weak coupling.

The last term of cascaded master equation is unique to the cascaded setting and breaks the symmetry between the emitters: it describes the process by which earlier emitter *i* radiates a photon that is then absorbed by later emitter *j*. It reduces to: $\dot{\rho} = -\frac{i}{\hbar}[H_{eff},\rho] + 2\gamma\sigma_-^{(2)}\rho\sigma_+^{(1)}$ (12) for two identical emitters coupled to common topological bath, where $H_{eff} = -i\hbar\frac{\gamma}{2}\left(\sigma_+^{(1)}\sigma_-^{(1)} + \sigma_+^{(2)}\sigma_-^{(2)}\right) - i\hbar\gamma\sigma_+^{(2)}\sigma_-^{(1)}$.

The far-field polarization distribution provides a sensitive measure of the emitters' mutual coherence and the chirality of their emission. We use $S_3$ stokes parameter which encode phase information and directly modulates the spin-angular momentum distribution of the emitted field of the non-trivial topological interference chirality to the far field. We visualize the polarization state of the emitted light from quantum emitters coupled to the chiral topological bath as a heterosystem (see Fig.4a) in two ways: (1) by looking at the Stokes parameters on the Poincaré sphere for various observation directions in *h* = 10 nm (Fig.4b-d), and (2) by mapping the spatial distribution of the circular polarization degree (Stokes $S_3$) in the far-field at different interaction regimes (Fig.e-j. in Fig. 4b, the polarization states demonstrated by the points (representing different far-field angles) in Poincaré sphere are highly clustered near the north pole of the sphere (red-colored points concentrated at $S_3 \approx$ +1) and highly clustered near the south pole of the sphere (blue-colored points concentrated at $S_3 \approx$ -1), red denoting σ⁺ and blue σ⁻. This indicates that the emission is almost right-circularly polarized – a signature of strong chiral coupling. We look at the same Poincaré sphere from +Z (the upper half) and -Z (the lower half) the coherent interreference pattern as the signature of emitters coherence are more visible (Fig. 4c,d). This coherence connects to collective jump operators of equation (11), $\sum_{i<j} 2\gamma_{ij}\sigma_-^{(j)}\rho\sigma_+^{(i)}$, which encode the inter-emitter coherence. Based on our observation, the degree of coherence across the aperture is high, yielding well-defined polarization states, we will discuss it more.

Fig.4e depicted spin-momentum locking is the non-trivial topological waveguide which the field is confined toward the interface, At the excitation point of topological interface $S_3$ becomes mixed due to the local interference of orthogonal fields and possibly multipolar radiation, when we put a gold NP as a quantum emitter in the nearfield of the topological interface ($h$=10 nm) (see Fig.4f), the quantum emitter couples strongly to the topological bath and absorb nonreciprocally and radiate it which act as a nanoantenna. The emitter radiation inherent the chiral property of the topological bath (predicted in Eq. (8)). We examine the radiation of two emitter located in small distance, $d$=200 nm from each other in the near field of the topological interface, the emitters act like a single superradiant source with a fixed handedness, facilitated by the near-field coupling and common topological edge-mode channel (see Fig.4g), which is evidence of coherent collective emission with preserved spin–momentum locking and predicted in equation (12).

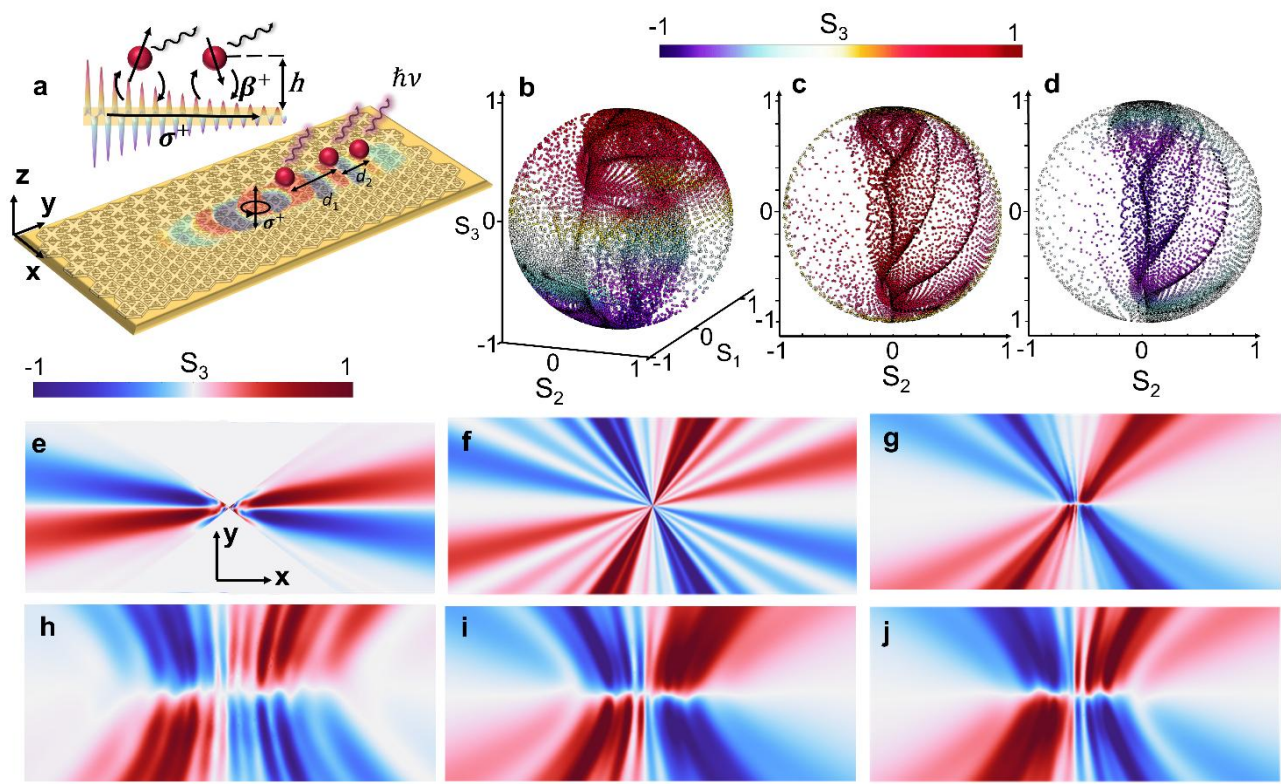

Figure 4: Chiral emitter coupling along a topological interface. **a**. Schematic of two (or more) 60-nm gold nanoparticle (red spheres) randomly placed along the topological edge of the plasmonic crystal at height $h$ (gold region with triangular hole lattice). The gold nano particles act as quantum emitter that couple into the guided helical edge mode (indicated by the colored evanescent field along edge and the $\sigma^+$ label for the propagation direction). A unidirectional coupling constant $\beta^+$ links an upstream emitter to a downstream emitter at random separation $d$, enabling chiral energy transfer. **b**. Poincaré sphere representation of far-field polarization states of light emitted from ten (10) coupled emitters located at height $h$=10 nm Each point on the sphere corresponds to the polarization (Stokes parameters $S_3$. Points are highly clustered near the north pole of the sphere (red-colored points concentrated at $S_3 \approx +1$) and near the south pole of the sphere (blue-colored points concentrated at $S_3 \approx -1$). **c,d**. The same Poincaré sphere at **b** from +Z (the upper half) and -Z (the lower half) the coherent interreference pattern as the signature of emitters coherence are more visible. **e**. depicted spin-momentum locking is the non-trivial topological waveguide which the field is confined toward the interface, at the excitation point of topological interface $S_3$ show mixed pattern, it may due to the local interference of orthogonal fields and multipolar radiation. **f**. A gold NP as a quantum emitter located in the nearfield of the topological interface ($h$=10 nm). The emitter radiation shows the same chirality of the topological interface mode. It is not propagation it is radiating, so it is not attached to interface. **g**. The radiation of two emitters located in small distance, $d$=200 nm from each other in the near field of the topological interface, the emitters act like a single superradiant source with a fixed handedness. **h**–**j**. Far-field polarization maps ($S_3$), at observation plane of 10 nm, 100 nm, 300 nm, respectively. Red regions denote predominantly $\sigma^+$ polarized intensity, blue regions $\sigma^-$. highly chiral, lobe-dominated emission at 10 nm in **h**. Transitions to more complex, fringe-patterned and less purely polarized emission at 300 nm in **j** demonstrate the transition from strong near-field chiral coupling to intermediate-/ weak-coupling regime.

In the second part, to further elucidate the coherence, we probe the far-field radiation resulting from 10 emitters randomly located along the interface, coupled via the edge as a function of their distance from the chiral bath (comparing $h$ = 10 nm, 100 nm, and 300 nm), which means we present far-field polarization maps for the three coupling strengths (See Fig. 4 (h-j)). We plot the spatial distribution of the circular polarization component $S_3$ for each case. The jump term $\sum_{i<j} 2\gamma_{ij}\, \sigma_-^{(j)} \rho \sigma_+^{(i)}$ in Lindblad master equation for cascade system (in Eq. (11)) builds coherent emission and interference pattern for $S_3$ polarization texture in far field.

For $h$=10 nm spaced emitters, see Figs. 4h, from the interference pattern, it is more pounced that from far-field polarization maps all emitters are excited in phase with the chiral reviser, coherently consistent with the $\sigma^+$ locked edge mode. At $h$=100 nm (Fig. 4i), interference fringes demonstrate a leakage is present at larger angles, a sign that slight phase differences allow a small portion of the radiation to have the opposite handedness in certain directions. Still, the central fringe is mostly red, showing the bias to $\sigma^+$ emission remains. At $h$=300 nm (Figs. 4j), multiple alternating red/blue fringes are clearly visible. The fringe spacing is narrower at this larger separation (angular separation scaling roughly inversely with $h$). The far-field maps thus vividly capture the gradual loss of global chirality and coherence as the emitter distance from the chiral bath increases, but even in $h$=300 nm case, within any given outgoing beam lobe, the polarization remains pure. This means the emission from each emitter is still highly coherent in itself (each emits $\sigma^+$ into the edge mode), but the relative phase delay between them leads to a spatially varying polarization after they scatter out.

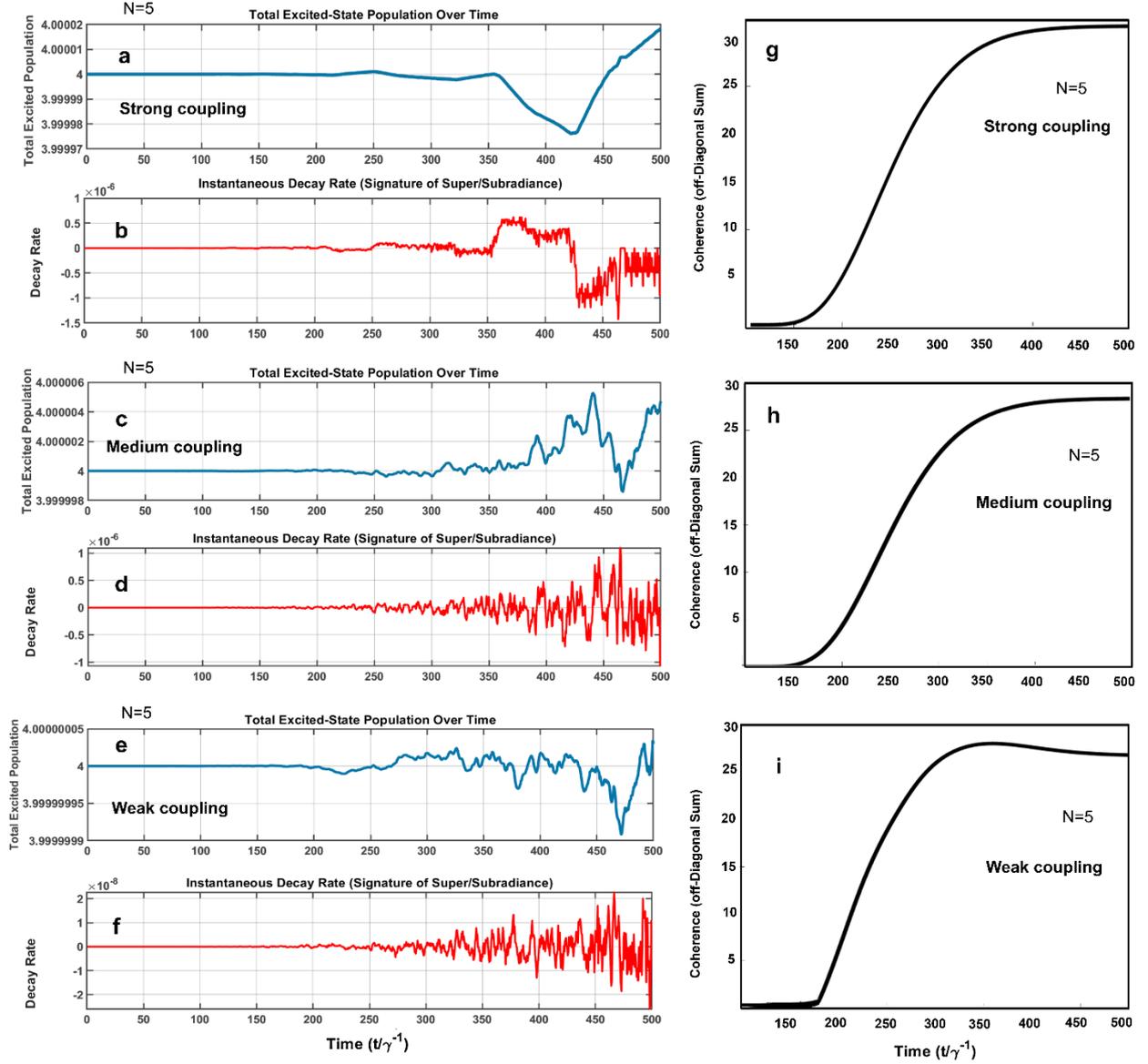

Figure 5: Temporal signatures of collective emission dynamics for different coupling regimes between quantum emitters and a chiral topological waveguide. Panels **a**–**f** display the time evolution of the total excited-state population (top) and instantaneous decay rate (bottom) for an ensemble of $N$=5 emitters coupled via a cascaded (unidirectional) master equation to a topological photonic reservoir. The emitter–bath interaction strength is varied to explore distinct regimes: **a**, **b**. Strong coupling: The system exhibits clear superradiant bursts, abrupt drops in excited-state population followed by long-lived subradiant plateaus, manifesting collective coherence and reabsorption via the chiral mode. **c**, **d**. Intermediate coupling: Both coherent and dissipative processes are evident. Fluctuations in decay rate indicate partial buildup of collective emission modes, with less pronounced subradiant trapping. **e**, **f**. Weak coupling: Despite being in the nominally incoherent regime, low-amplitude oscillations in the decay rate persist, implying residual phase coherence mediated by the structured chiral bath. The slow decay and non-monotonic dynamics suggest that the topological reservoir supports non-trivial interference effects even under weak excitation. In all regimes, the instantaneous decay rate acts as a diagnostic of super- and subradiant behavior, while the excited-state population traces the redistribution of quantum excitation within the emitter ensemble. **g**–**i**: The corresponding growth of total off-diagonal coherence further confirms the robustness of collective phase locking across all regimes. These results demonstrate that topologically protected edge modes not only enable non-reciprocal energy transfer but also preserve coherence even in weak coupling, challenging conventional expectations of incoherence in open quantum systems.

**Discussions**

Our findings establish a robust and tunable platform for engineering collective light–matter interactions mediated by a topologically protected chiral photonic environment. By integrating quantum emitters near a topological plasmonic interface, we demonstrate, both theoretically and numerically, that the emergence of long-range quantum coherence and cooperative emission phenomena such as super- and subradiance is not confined to the strong-coupling regime. Instead, the spatially structured and unidirectional nature of the chiral edge mode enables phase-coherent dynamics that persist even in the weak-coupling limit, beyond the predictive reach of conventional waveguide QED models. In traditional open quantum systems, weak coupling to the environment typically induces rapid decoherence, primarily due to inefficient reabsorption and the stochastic nature of phase diffusion. However, in our system, this paradigm is fundamentally altered by the presence of topologically protected chiral edge states. These states not only guarantee unidirectional propagation but also exhibit intrinsic robustness against disorder, backscattering, and local perturbations. Consequently, even quantum emitters weakly coupled to the photonic reservoir experience phase-stable, directional interactions, which are typically absent in non-topological platforms. This resilience manifests through non-reciprocal excitation exchange and the suppression of environment-induced noise. In our master equation formalism, this is captured by two key ingredients: (i) the inclusion of directionally biased (cascaded) coherent interactions in the effective Hamiltonian, and (ii) a dephasing superoperator $\mathcal{L}_\varphi[\rho]$, representing environmental noise that typically erodes quantum coherence. Despite the presence of $\mathcal{L}_\varphi$, we observe that coherence and quantum interference persist, underscoring the exceptional phase-preserving nature of the chiral edge channel. Far-field polarization maps, particularly interference structures in the Stokes parameter $S_3$ provide direct signatures of emitter coherence. These spatial patterns correspond tightly to the temporal quantum dynamics extracted from cascaded Lindblad simulations, in which the instantaneous decay rate serves as a dynamical marker of collective emission. In the strong-coupling regime, we observe pronounced superradiant bursts followed by extended subradiant plateaus, consistent with population trapping in decoherence-free subspaces. Remarkably, even in the weak-coupling regime, low-amplitude modulations in the decay rate and coherent $S_3$ interference fringes persist, confirming the role of the chiral interface in protecting quantum correlations. This behavior is vividly illustrated in Figure 5, where time-domain simulations show collective decay dynamics across various coupling regimes. The persistence of structured $S_3$ polarization textures and temporal coherence at emitter separations exceeding 300 nm reveals that the topological waveguide supports long-range dipole–dipole interactions, well beyond the near-field regime. These effects are inherently non-Markovian, enabled by the unidirectional and phase-preserving nature of the chiral photonic channel. Together, Figures 4 and 5 offer robust cross-validation between spatial polarization topology and time-resolved coherence. The observed phenomena signify more than a passive suppression of decoherence, they represent a nontrivial light–matter coupling scenario where topologically mediated directionality and phase locking facilitate the formation of entangled emitter–photon states. The ability of an excitation to propagate directionally over long distances with preserved phase information not only connects distant emitters but also opens new opportunities for controlling the decoherence, realizing scalable entanglement networks, nonreciprocal quantum logic, and phase-sensitive quantum photonic devices beyond the decoherence limitations. The implications of our findings are not limited to the use of gold nanoparticles as quantum emitters. More broadly, this work opens a pathway to control how quasiparticles, such as excitons, phonons, magnons, polaritons, and plexcitons, interact in heterostructured systems. In particular, systems based on two-dimensional materials with engineered layers, topological architectures, and metamaterials host novel quasiparticle states that could benefit from the coherence-preserving principles demonstrated here. Our platform offers a strategy to exploit or stabilize quantum coherence in such environments, enabling the generation of localized, long-lived quantum states in solid-state systems.

**Acknowledgements**

This work was supported by the funding of the Early Career Award of the Priority Research Area Kiel Nano Surface and Interface Science (KiNSIS), Kiel University.